\documentclass[twocolumn,aps,prl,superscriptaddress,amssymb,showpacs]{revtex4}
\usepackage[dvips,final]{graphicx}
\begin{document}
\title{Negative differential resistance in
normal narrow bands - superconducting junctions with Andreev
reflection} 
\author{N. Garc\'ia}\email{nicolas.garcia@fsp.csic.es}
\affiliation{Division of Superconductivity and Magnetism, Institut
f\"ur Experimentelle Physik II, Universit\"{a}t Leipzig,
Linn\'{e}stra{\ss}e 5, D-04103 Leipzig, Germany}
\affiliation{Laboratorio de F\'isica de Sistemas Peque\~nos y
Nanotecnolog\'ia,
 Consejo Superior de Investigaciones Cient\'ificas, E-28006 Madrid, Spain}
\author{P. Esquinazi}\email{esquin@physik.uni-leipzig.de}
\affiliation{Division of Superconductivity and Magnetism, Institut
f\"ur Experimentelle Physik II, Universit\"{a}t Leipzig,
Linn\'{e}stra{\ss}e 5, D-04103 Leipzig, Germany}

\begin{abstract}
We have calculated current-voltage characteristic curves for
normal-superconducting junctions with Andreev reflections and
different types of electronic bands. We found that when the normal
band is narrow, of the order of the superconducting energy gap, a
negative differential resistance appears at a voltage of the order
of the band width plus the gap value. In case two bands contribute
to the total current  the conductance can be smaller than unity at
voltages above the gap value. Our simulations may provide an
answer to different experimental data of the literature that were
not yet understood.
\end{abstract}  \maketitle

The measurement of the current-voltage $(I-U)$ characteristic curves
of junctions made from similar or different materials separated by an
insulating layer is one of the usual transport methods to obtain
information on the electronic band structures near the Fermi level
$E_F$. If one of the materials of the junctions is a superconductor,
the $I-U$ curve provides a direct measurement of the energy gap
$\Delta$, for example. In this case the voltage region smaller than
$\Delta/e$ ($e$ is the electronic charge) the conductance $G$ is in
general  much smaller than $G_N$, the value obtained at $U \gg
\Delta/e$. In case one has a plain normal-superconducting junction
without any insulating intermediate layer, the conductance $G(U <
\Delta/e) \sim 2 G(U \gg \Delta/e) = 2 G_N$, a rather surprising
phenomenon taking into account that for a normal-normal metals
junctions the normalized conductance $G/G_N = 1$ at all voltages.

Andreev studied theoretically the heat transport through a
normal-superconducting plain junction and obtained the factors 2 and
1 for the normalized conductance in these two voltage regions
\cite{and64}. Blonder, Tinkham and Klapwijk (BTK) \cite{blo82}
studied the reflection of an electron at this kind of junctions and
obtained a similar solution for the conductance, i.e. it decreases
from a factor 2 to 1 when the voltage grows from zero to values above
the energy gap of the superconducting part of the junction. The
physical explanation for the value of 2 at $U < \Delta/e$ is related
to the reflection and transmission of the electron of the normal part
into the superconducting one. The electron current from the normal
part transforms in a Cooper pair current in the superconductor and a
hole reflects back into the normal part. Using a different approach
Garcia, Flores and Guinea \cite{gar88} studied a similar phenomenon
observed in scanning tunneling microscopy and obtained the same
equation as in Ref.~\cite{blo82}.

All the $I-U$ curves discussed above were obtained for the case of
ballistic transmittance between a normal metal and a
superconductor. This approach has been extended for the case of
ferromagnets with a certain spin polarization and for a diffusive
regime in a recent publication \cite{maz01}.  In the last years
there have been several reports
\cite{sou98,wei98,sou99,rou05,pan05,str01,panprb05,yun07} showing
experimental data with conductance  values less than its normal
state value $G_N$ at $U
> \Delta/e$. Although some modeling for this behavior
 on the basis of non-ideal interfaces, proximity effects and energy losses
 were proposed, the generality of this
 behavior observed in different kind of samples suggests that
 other explanation may be needed.

\begin{figure}[]
\begin{center}
\includegraphics[width=1.1\columnwidth]{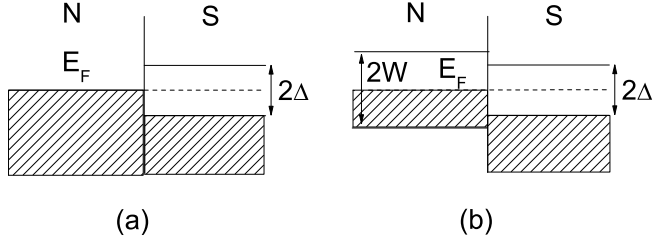}
\caption[]{Energy band structures at the normal (N) - superconducting
(S) junctions. We used two main models for the normal part of the
junction: (a) Usual N-S junction. (b) Junction between a
superconductor and a semiconducting material with a band width $2W$.
Both graphs are valid at $T = 0$~K and at no applied bias voltage
$U$.} \label{theo}
\end{center}
\end{figure}

Assuming two different main cases for the normal-superconducting
junctions, see Fig.~\ref{theo}, we calculate the $I-U$
characteristic curves and conductance as a function of bias
voltage $U$. Our aim is to show in which cases the conductance
gets smaller than unity (in reduced scale, i.e. $G/G_N$) at
voltage values above the gap implying the contribution of a
negative differential resistance to the total current. We show
below that this can be observed when the band width $W$ of the
normal material is of the order of the energy gap of the
superconducting material, what we call a normal narrow band. The
conductance exhibits a negative differential resistance with a
maximum absolute value at $\Delta < eU \lesssim \Delta + W$,
specially in the two dimensional case and it tends to zero for
voltages $eU > W+\Delta$.

We assume that the junction is composed by two different materials
one of them a superconductor with a energy gap $\Delta$. Following
Refs.~\cite{blo82,gar88} the Andreev conductance at zero temperature
(in units of quantum of conductance)  $G(U)$ is given by

\begin{equation}\label{GU}
G(U) =     1 - \frac{|(1-a^2)|^2 (1- |T|)}{|1-a^2(1-|T|)|^2}
      + \frac{a^2|T|^2}{|1-a^2(1-|T|)|^2} \,,
\end{equation}
which depends on the transmittivity $T$ in the {\it normal state}; $a
= (U/\Delta) - [(U/\Delta)^2 - 1]^{0.5}$.

The current $I$ as a function of the voltage $U$ at the junction can
be calculated taking into account that  it is controlled by Cooper
pairs Andreev currents if $eU<\Delta$ \cite{and64}. However at higher
$U$ the current is controlled by quasiparticles and for $U>7\Delta/e$
we have practically conduction between normal materials
\cite{and64,blo82,gar88}. The solution for the current $I$ is
\begin{eqnarray}
        I(U) &=&  B \cdot A \int_0^U G(U')  dU' \,, \label{I}
\end{eqnarray}
where $A$ is a constant that depends on the junction geometry and
on the integration average on $\alpha$, the angle that the
incoming particles form to the interface and $B = N(0) e v_F$,
with $N(0)$ the density of states at  $E_F$ and $v_F$ the Fermi
velocity. This integral corresponds to ballistic 3D case for a
system in which the occupied normal band is much broader than
$\Delta$, i.e. $I_{W \gg \Delta}$. The transmittivity can be
expressed as $T = 1/(1 + Z_0^2)$, where $Z_0$ was defined in
\cite{blo82}. Notice that $Z_0 = 0$ and $\infty$ means $T = 1$ and
0.

\begin{figure}[]
\begin{center}
\includegraphics[width=1.1\columnwidth]{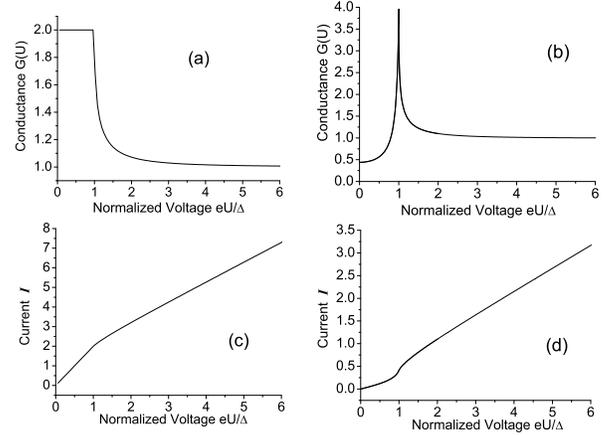}
\caption[]{(a) Theoretical conductance $G$ vs. normalized voltage
$eU/\Delta$ for $T = 1$ calculated using Eq.~(\protect\ref{GU}) (the
case of Fig.~\ref{theo}(a)). (b) The same as in (a) but for $T = 0.5$
and normalized to its value at $eU \gg \Delta$. (c) Total current $I$
calculated with (\protect\ref{I}) for $T =1$ and (d) $T = 0.5$.}
\label{fig2}
\end{center}
\end{figure}

For the case of Fig.~\ref{theo}(a) the conductance (calculated
assuming a transmittivity $T = 1$ , see Eq.~(\ref{GU})) has the known
value of 2 at $U \le \Delta$ and reduces gradually to 1 at voltages
above $\Delta$, see Fig.~\ref{fig2}(a). The total current $I$ through
the junction is calculated following Eq.~(\ref{I}) and is depicted in
Fig.~\ref{fig2}(c). In case we have $T = 0.5$ we get the results
shown in Fig.~\ref{fig2}(b,d) for the conductance and current. With
exception of the curves in Fig.~6, all others are normalized to their
normal-state values.

In the case we have a normal metal with a narrow band width $W \sim
\Delta$, then the situation changes  and the equation for the current
is given by

\begin{equation}\label{I23D}
    I_{W \sim \Delta} = A m^\star \int_0^U (W-eU')^b (E_F + eU')^b G(U')  dU' \,,
\end{equation}
where the parameter $b = 0.5, 0, -0.5$ corresponds to the 3D, 2D and
1D case and the parameter $A$ has the same meaning as in
Eq.~(\ref{I}) and $m^\star$ is the effective mass of the carriers.
The correction term $(E_F + eU')^b$ can be in general neglected if
$E_F \gg eU$. The theoretical results below are calculated at zero
temperature.

\begin{figure}[]
\begin{center}
\includegraphics[width=1.1\columnwidth]{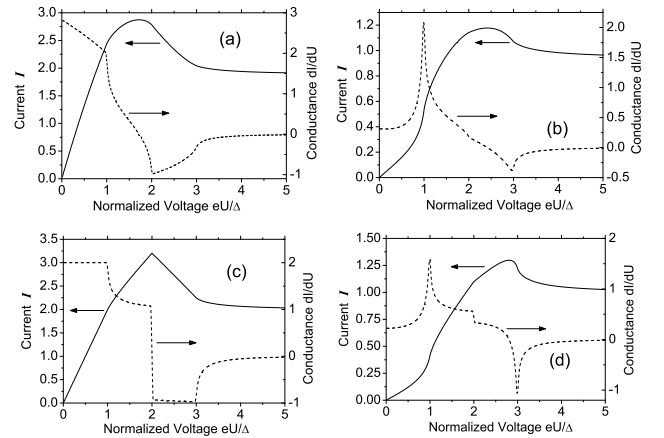}
\caption[]{Continuous lines: Total current $I$ vs. normalized voltage
$eU/\Delta$ for the case of a normal part with a narrow band $W =
2\Delta$ for $T = 1$ and $0.5$ for the 3D (a,b) and 2D case (c,d).
Dashed lines represent the calculated conductance $dI/dU$. These
calculations correspond to the case of Fig.~\ref{theo}(b).}
\label{fig4D}
\end{center}
\end{figure}

If we have a narrow band in energy with its minima at $U = 0$ near
the superconducting gap, the integral to estimate the current is like
in Eq.~(\ref{I23D}) but with the bottom of the band that depends on
the assumed band width $2W$ and therefore of the order of $\Delta$.
Note that in this case the current will not be symmetric for positive
and negative voltages. For positive voltages Fig.~\ref{fig4D} shows
the current for the case of a conductor with a narrow band (see
Fig.~\ref{theo}(b)) filled up to $W = 2 \Delta$ in the 3D and 2D
cases for two different transmittivities. Note that in both 3D and 2D
cases there appears a negative differential resistance region at $2
\lesssim eU/\Delta \lesssim 3$ for $T = 1$ and at slightly higher
voltage values for $T = 0.5$, see Fig.~\ref{fig4D}. The behavior of
the conductance $dI/dU$ depends on the transmittivity value as well
as the dimensionality of the normal band, see dashed lines in
Fig.~\ref{fig4D}. In case the voltage is negative we will obtain the
same curve if the Fermi energy were at the middle of the band.
Otherwise we need to do the calculations for the specific case.

\begin{figure}[]
\begin{center}
\includegraphics[width=1.1\columnwidth]{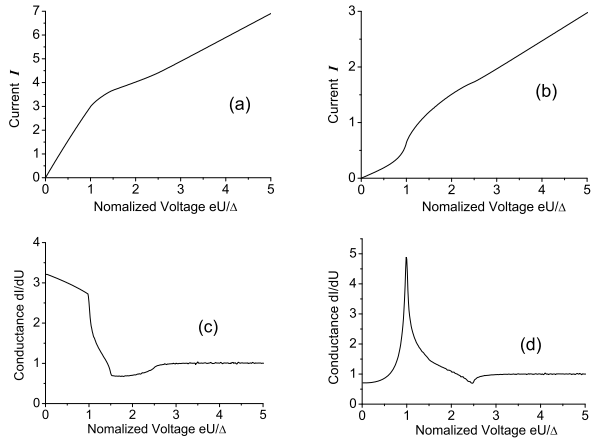}
\caption[]{(a) Total current $I$ vs. normalized voltage $eU/\Delta$
for $T = 1$ for the case of a normal part having two materials
touching in parallel the superconducting part, one with $W \gg
\Delta$ and a second with $W = 1.5~\Delta$, calculated using $I =
I_{W\gg\Delta} + 0.5 I_{W \sim \Delta}$ (Eq.~(\protect\ref{I}) and
Eq.~(\protect\ref{I23D})) for the 3D case. (b) The same as in (a) but
for $T = 0.5$. (c) and (d) show the corresponding conductances
$dI/dU$. In (d) the conductance is normalized by its value at $eU \gg
\Delta$.} \label{fig3}
\end{center}
\end{figure}

\begin{figure}[]
\begin{center}
\includegraphics[width=1.1\columnwidth]{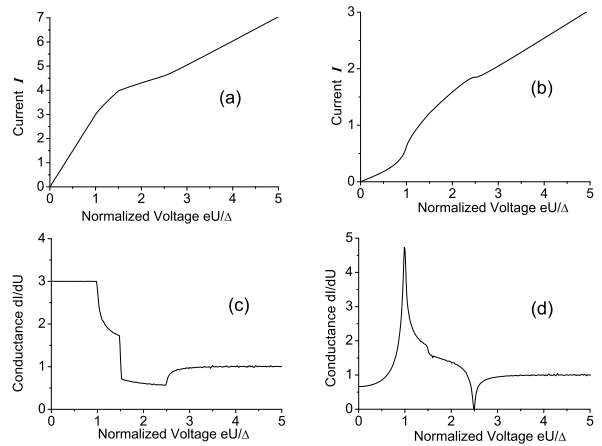}
\caption[]{(a) Total current $I$ vs. normalized voltage $eU/\Delta$
for the case of a normal part having two materials touching in
parallel the superconducting part, one with $W \gg \Delta$  and a
second one with $W = 1.5~\Delta$ for $T = \exp(-1)$ for the 2D case.
(b) The same as in (a) but for $T = \exp(-0.5)$. (c) and (d) show the
corresponding conductances $dI/dU$ normalized by its value at $eU \gg
\Delta$.} \label{fig5}
\end{center}
\end{figure}

\begin{figure}[]
\begin{center}
\includegraphics[width=1.1\columnwidth]{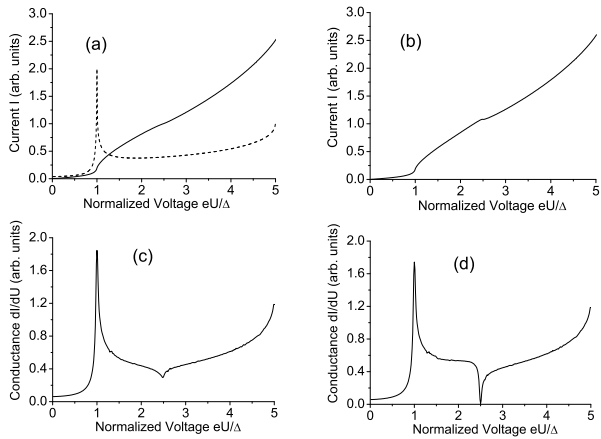}
\caption[]{Total current vs. normalized voltage for a normal
conducting part with two parallel contributions with $W \gg \Delta$
and $W = 1.5 \Delta$ with a transmissivity $T = \exp(-(2 - 0.4
U)^{0.5})$ for the 3D case (a) and 2D case (b). The dashed line in
(a) is the conductance $G$ following Eq.~(\protect\ref{GU}). The
figures in (c) and (d) show the corresponding conductances $dI/dU$.}
\label{fig7}
\end{center}
\end{figure}

One could also treat the case in which the narrow band material is
present in parallel to the larger band one,  both contributing in
parallel to the current through the junction. This is an interesting
case that may occur at the interfaces of the contacts.  When the two
materials, the normal and superconducting one, are put together
narrow bands at the Fermi energy may appear upon materials used,
junction geometry and quality. The current and the conductance are
presented in Figures \ref{fig3} to \ref{fig7} for different cases.

In the particular cases treated below we have reduced the
transmittivity of the narrow band respect to the large one by a
factor of two. This assumption means that the effective mass of
the carriers in the smaller band is half of that in the larger
one. Nevertheless, we will see that the negative differential
resistance contribution from the narrow band influences the total
conductance. We note that qualitatively similar characteristic
curves as we describe below were observed in many experiments but
in general their meaning was not discussed. The cases we discuss
below can really
 happen because these narrow bands can be formed at the contacts between
 the normal and superconducting parts.
 Figure~\ref{fig3} shows the calculated total current $I$ given by
the sum of two normal conductors in parallel, one with a large
band width $W \gg \Delta$ and the other with $W = 1.5 \Delta$, for
$T = 1$ (a) and $T = 0.5$ (b). The figures (c) and (d) show the
calculated conductances $dI/dU$. Figure~\ref{fig5} shows the
results for the 2D case and for $T = \exp(-1)$ or $T =
\exp(-0.5)$.

We describe now the last case when the small and large bands
contribute to the current but their carriers have to overcome a
barrier of height $\phi$ existing between them and the
superconductor. If $\phi$ is not very large then it will allow more
current as the potential $U$ increases. An example can be seen in
Ref.~\cite{yun07}. Figure~\ref{fig7} shows the results of the
calculations for two bands using $T = \exp(-(2 - 0.4U)^{0.5})$. It
can be seen that at low voltages the conductance is smaller than at
high voltages with an intermediate region where it decreases due to
the negative resistance contribution from the narrow band. This case
is different from the previous cases, see for example
Figs.~\ref{fig3} and \ref{fig5}, because there the bands just
contribute with different transmittivities but these do not depend on
$U$, i.e. there is no tunnel barrier that change the conductivity
appreciably. The increase of the conductance at large $U$ values, as
is the case of Fig.~\ref{fig7}, may occur because when a contact is
formed it could have some oxide that acts as a potential barrier.

In conclusion, in this paper we have discussed the contribution of
different types of bands of normal--superconducting junctions with
Andreev reflections. The results show: (1) For a single normal wide
band $W \gg \Delta$ the conductance $G$ and the current vs. applied
voltage behave as expected, i.e. $G = 2$ for $eU \le \Delta$ and
tends to 1 for $eU > \Delta$. (2) For a junction with a normal
narrow-band width $2W \sim \Delta$ and at $eU > W + \Delta$ the
current tends to a constant and the conductance to zero. At
intermediate voltages we find a region of negative differential
resistance (current decreases with voltage) as shown in
Fig.~\ref{fig4D}. (3) As two bands, one with a large and the other
with a narrow band width, contribute in parallel, a variety of cases
appear. In any of those discussed here the negative differential
resistance part coming from the narrow band has a clear influence on
the expected $I-U$ characteristics. (4) Finally, if we assume a not
very high potential barrier between the normal and superconducting
parts, the conductance does not saturate but steadily increases at
high voltages $eU > W+ \Delta$. These results indicate that two
electronic bands at the Fermi level can complicate the form of the
characteristic $I-U$ curves in real junctions. A comparison of our
results with the available experimental data
\cite{sou98,wei98,sou99,rou05,pan05,str01,panprb05,yun07} indicates
that our model should be useful to understand experimental results.

For wide band ferromagnetic materials and ballistic transport (which
produces minima for bias voltages smaller than the energy gap)
further extensions of the BTK model were reported in
Ref.~\cite{maz01,xia02}, approaches that can be further developed
including the contributions of narrow bands. A narrow band effect can
be also treated taking into account a proximity effect \cite{str01}.
However, if the conductance minima are at higher voltages than the
equivalent to the energy gap (as e.g. the case of Fe/Ta
\cite{sou98}), the proximity effect  can unlikely explain it, but it
is a clear sign for a narrow band effect. These possibilities have
not been yet discussed in any of the existing experimental papers and
stress the possible existence of interesting physics at contacts and
their interfaces.

We gratefully acknowledge the collaboration and support of J.
Barzola-Quiquia. This work was support by: DAAD (Grant No.
D/07/13369, ``Acciones Integradas Hispano-Alemanas"), SFB~762
``Funkionalit\"at Oxidischer Grenzfl\"achen" and Spanish CICyT. N.G.
is supported by the Leibniz professor fellowship of the University of
Leipzig.


\end{document}